\documentclass[11pt]{article}
\newcommand{\Comment}[1]{{}}
\usepackage{amsfonts,amsthm,amsmath,amssymb,slashed}
\usepackage[textwidth = 430 pt, textheight = 630 pt]{geometry}
\usepackage{color}

\Comment{\usepackage{color}
\definecolor{MyDarkBlue}{rgb}{0.15,0.15,0.45}
\usepackage[linktocpage=true]{hyperref}
\hypersetup{
colorlinks=true,
citecolor=MyDarkBlue,
linkcolor=MyDarkBlue,
urlcolor=MyDarkBlue,
pdfauthor={Daniel W.F. Alves, Carlos Hoyos, Horatiu Nastase and Jacob Sonnenschein},
pdftitle={The title},
pdfsubject={hep-th}
}

\usepackage[numbers,sort&compress]{natbib}
\usepackage{hypernat}}
\usepackage{graphicx}
\usepackage{cite}

\newcommand\ignore[1]{}
\def\one{{\,\hbox{1\kern-.8mm l}}}

\def\a{\alpha}\def\b{\beta}

\def\d{\partial}

\newcommand{\Cset}{{\,\,{{{^{_{\pmb{\mid}}}}\kern-.45em{\mathrm C}}}}}

\newcommand{\be}{\begin{equation}}
\newcommand{\bea}{\begin{eqnarray}}

\newcommand{\ee}{\end{equation}}
\newcommand{\eea}{\end{eqnarray}}

\begin{document}

\renewcommand{\thefootnote}{\fnsymbol{footnote}}

\makeatletter
\@addtoreset{equation}{section}
\makeatother
\renewcommand{\theequation}{\thesection.\arabic{equation}}

\rightline{}
\rightline{}




\begin{center}
{\LARGE \bf{\sc Hopfion solutions in gravity and a null fluid/gravity conjecture}}
\end{center} 
 \vspace{1truecm}
\thispagestyle{empty} \centerline{
{\large \bf {\sc Daniel W.F. Alves${}^{a}$}}\footnote{E-mail address: \Comment{\href{mailto:dwfalves@ift.unesp.br}}{\tt dwfalves@ift.unesp.br}}
{\bf{\sc and}}
{\large \bf {\sc Horatiu Nastase${}^{a}$}}\footnote{E-mail address: \Comment{\href{mailto:horatiu.nastase@unesp.br}}{\tt horatiu.nastase@unesp.br}}
                                                        }

\vspace{.5cm}


\centerline{{\it ${}^a$Instituto de F\'{i}sica Te\'{o}rica, UNESP-Universidade Estadual Paulista}} 
\centerline{{\it R. Dr. Bento T. Ferraz 271, Bl. II, Sao Paulo 01140-070, SP, Brazil}}

\vspace{1truecm}

\thispagestyle{empty}

\centerline{\sc Abstract}

\vspace{.4truecm}

\begin{center}
\begin{minipage}[c]{380pt}
{\noindent 
We conjecture an extension of the fluid/gravity correspondence to the null pressureless fluid case via gravitational shockwave solutions, and 
use it to propose an embedding of the fluid Hopfion in gravity. A nonlinear gravitational "helicity" is also proposed, analogous with the helicity of 
electromagnetism and fluid dynamics.

}
\end{minipage}
\end{center}

\vspace{.5cm}

\setcounter{page}{0}
\setcounter{tocdepth}{2}

\newpage

\renewcommand{\thefootnote}{\arabic{footnote}}
\setcounter{footnote}{0}

\linespread{1.1}
\parskip 4pt



\section{Introduction}

Nonlinear field theories sometimes have soliton solutions characterized by some topological charge. It is less known that Maxwell's electromagnetism, a free theory, 
has also nontrivial solutions characterized by a topological index, the Hopf index, and moreover a linking number, the "Hopfion" solutions. They were found rather late, 
by Ra\~nada in \cite{Ranada:1989wc,ranada1990knotted}, based on earlier work by Trautman \cite{Trautman:1977im}. These solutions satisfy a null condition $\vec{F}^2=0$, 
where $\vec{F}=\vec{E}+i\vec{B}$ is the Riemann-Silberstein vector, so $\vec{E}\cdot \vec{B}=\vec{E}^2-\vec{B}^2=0$, and are characterized by some complex scalar functions
$\a$ and $\b$. They are characterized by "helicities" $H_{ee},H_{mm},
H_{em}$ and $H_{me}$ (where $e$ stands for electric and $m$ for magnetic), which are written as spatial integrals of Chern-Simons or BF type terms $\int d^3x \epsilon^{ijk}
A_i\d_j B_k$, which are conserved (time independent) for null configurations. They also have a nonzero Hopf index and linking number for $\vec{E}$ with $\vec{B}$, but we 
will not discuss that in the following. One can also obtain more general "$(p,q)$-knotted" 
solutions by using the transformations $\a\rightarrow \a^p$, $\b\rightarrow \b^q$, where $\a$ and $\b$ define 
the solution through $\vec{F}=\vec{\nabla}\a\times\vec{\nabla}\b$
\cite{Kedia:2013bw,besieris2009hopf,Hoyos:2015bxa}. For a more complete review, see \cite{arrayas2016knots}.

One can write knotted solutions in fluid dynamics as well. Despite the fact that knot theory was developed by Lord Kelvin based on fluid knots (after earlier work by 
Helmholtz in 1858), explicit solutions were lacking until rather recently. This was started by the observation of Moffatt \cite{moffatt1969degree} that one can 
write a fluid helicity $H_v=\int d^3x \epsilon^{ijk}v_i \d_j v_k$, similar to $H_{mm}$ of electromagnetism, where the fluid velocity $\vec{v}$ is analog of the vector potential $\vec{A}$ of electromagnetism, 
and writing some solutions with $H_v\neq 0$ in some very special cases, further studied in \cite{moffatt1992helicity,moffatt1992helicity2}. Further solutions were 
found in \cite{ValarMorgulis,crowdy_2004, Ifidon2015216}, and initial conditions with nonzero $H_v$ were described in \cite{kuznetsov1980topological}.
For our purposes however, the relevant solutions were found in \cite{Alves:2017ggb,Alves:2017zjt}, 
where it was found that electromagnetism can be 
written as a kind of null pressureless fluid ($P=0, u^\mu u_\mu=0$), allowing us to write a null fluid Hopfion solution for fluid dynamics. One also obtained a ``nonrelativistic fluid Hopfion'' in 2+1 dimensions, by a certain dimensional reduction 
procedure.

One can ask whether we can also write a Hopfion solution for gravity. On the other hand, it is known that there is a fluid/gravity correspondence
(see  \cite{Hubeny:2011hd,Rangamani:2009xk,Eling:2010vr} for a review)  relating fluid dynamics with gravity near the horizon of modified black hole solutions. 
Given the previous discussion, it seems like the correct way to embed the Hopfion into gravity, via the null 
fluid Hopfion solution (or in 2+1 dimensions, via the nonrelativistic fluid Hopfion). One possible problem is that 
the original Hopfion was a solution to a free theory, whereas gravity is nonlinear; but that was true also in fluid dynamics, and the solution there was to focus on the null 
pressureless sector which is simpler. It will turn out that the same is true for gravity, and the null sector we need to consider corresponds to a set of gravitational shockwaves, 
which are known to be simpler, and to linearize gravity in some cases. However, it also turns out that the fluid/gravity correspondence was not defined for these null fluids, 
corresponding to shockwaves, so we define it (as a somewhat straightforward generalization) here. It is still conjectural, since it turns out to be even more difficult than in the 
original case to check that the gravitational equations are satisfied if the null fluid equations are, but we will give some plausibility arguments. 

Also, in order for the the equivalent of the Hopfion solution to exist, it seems reasonable that there should be an analog of the helicities of electromagnetism and fluid 
dynamics. We show indeed that this is the case, and we can define a nonlinear quantity that can take the role of gravitational helicity, and show the condition for it to be 
conserved (time-independent). It reduces to the eletromagnetic helicity in some particular (and linearized) case.

The paper is organized as follows. In section 2 we review the electromagnetic and null fluid Hopfion solutions. In section 3 we describe our proposal for a 
nonlinear helicity in gravity, and we argue it based on the similarities between linearized gravity and electromagnetism. In section 4 we describe our conjecture for a null 
fluid/gravity correspondence, and using it, we embed the null fluid Hopfion in gravity, showing that it has a nontrivial gravitational helicity, and in section 5 we conclude.

\section{Review: electromagnetic and null fluid Hopfions}

The basic "Hopfion" solution of electromagnetism is simplest to describe in the construction of Bateman. Consider the 
Riemann-Silberstein vector
\be
\vec{F}=\vec{E}+i\vec{B}\;,
\ee
in terms of which the Maxwell equations become
\be
\vec{\nabla}\times \vec{F}=i\frac{\d}{\d t}\vec{F};\;\;\; \vec{\nabla}\cdot \vec{F}=0\;,
\ee
and solve automatically the second equation by the ansatz 
\be
\vec{F}=\vec{\nabla}\a \times \vec{\nabla}\b.
\ee
Then the resulting Maxwell equation reduces, on the ansatz, to 
\be\label{Batequsol}
i (\d_t \a \vec{\nabla}\b-\d_t\b\vec{\nabla}\a)=\vec{F}=\vec{\nabla}\a\times \vec{\nabla}\b.
\ee
In the case that these equations are satisfied, the vector $\vec{F}$ must be null, i.e.,
\be
\vec{F}^2=0\Rightarrow \vec{E}^2-\vec{B}^2=0,\;\;\;\; \vec{E}\cdot\vec{B}=0.
\ee
Then the basic Hopfion solution is given by 
\bea
\a&=&\frac{A-1+iz}{A+it};\cr
\b&=&\frac{x-iy}{A+it}\cr
A&=& \frac{1}{2}(x^2+y^2+z^2-t^2+1).\label{emhopfion}
\eea
and we can check that it satisfies (\ref{Batequsol}), and is null, $\vec{F}^2=0$. The explicit formulas for $\vec{E},\vec{B}$ can 
be found for instance in  \cite{Alves:2017zjt,Alves:2017ggb}, though they are quite long, so we will not reproduce them here. 
The explicit formulas for the gauge potentials $\Phi$ and $\vec{A}$ are not available, but could be obtained from them.

In the absence of sources, one has electric-magnetic duality, relating $\vec{E}$ and $\vec{B}$ which means we can define also
a vector $\vec{C}$ by (in the gauge $A_0=-\Phi=0$)
\be
\vec{E}=\vec{\nabla}\times \vec{C}=-\d_t \vec{A};\;\;\; \vec{B}=\vec{\nabla}\times \vec{A}=-\d_t \vec{C}.\label{potentials}
\ee

One can then define the "helicities" of sourceless electromagnetism, quasi-topological quantities defined as spatial Chern-Simons 
forms made up from $\vec{A}$ and $\vec{C}$,
\bea
H_{ee}&=&\int d^3x \vec{C}\cdot \vec{E}=\int d^3x \vec{C}\cdot \vec{\nabla}\times \vec{C}=\int d^3x \epsilon^{ijk}C_i\d_j C_k\cr
H_{mm}&=&\int d^3x \vec{A}\cdot \vec{B}=\int d^3x\epsilon^{ijk}A_i\d_j A_k\cr
H_{em}&=&\int d^3x \vec{C}\cdot\vec{B}=\int d^3x \epsilon^{ijk}C_i\d_j A_k\cr
H_{me}&=&\int d^3x \vec{A}\cdot\vec{E}=\int d^3x \epsilon^{ijk}A_i\d_k C_k\;,
\eea
and we can show that these are time independent if $\vec{E}\cdot \vec{B}=0=\vec{E}^2-\vec{B}^2$. In particular, for 
the null Hopfion solution, this is true, and the helicities are conserved, and are moreover nontrivial: one finds that 
for the Hopfion $H_{ee}=H_{mm}\neq 0$ and $H_{em}=H_{me}=0$.

As seen in \cite{Alves:2017zjt,Alves:2017ggb}, we can map electromagnetism to a null pressureless fluid, with 
\be
T_{\mu\nu}=\rho u_\mu u_\nu\;,
\ee
and $u^\mu u_\nu=0$, which interestingly gives the same continuity equation and Euler equation as the nonrelativistic ideal 
fluid, by the map 
\be
\rho\leftrightarrow \frac{1}{2}(\vec{E}^2+\vec{B}^2);\;\;\;\;
v_i\leftrightarrow\frac{[\vec{E}\times \vec{B}]_i}{\frac{1}{2}(\vec{E}^2+\vec{B}^2)}.\label{velmap}
\ee
For a traceless electromagnetic 
energy-momentum tensor, ${T^\mu}_\mu=0$, this gives a null fluid, $\vec{v}^2=1$, which then in turn implies
a null electromagnetism, $\vec{F}^2=0$. 

One can then use the Hopfion solution to find a corresponding "null fluid Hopfion" solution, which then has energy density
\be
\rho = \frac{16 \left((t-z)^2+x^2+y^2+1\right)^2}{\left(t^4-2 t^2
   \left(x^2+y^2+z^2-1\right)+\left(x^2+y^2+z^2+1\right)^2\right)^3}
\ee
and velocity
\bea
v_x&=&\frac{2(y+x(t-z))}{1+x^2+y^2+(t-z)^2},\cr 
v_y&=&\frac{-2(x-y(t-z))}{1+x^2+y^2+(t-z)^2},\cr
v_z^2&=&1-v_x^2-v_y^2.\label{velhopf}
\eea

But we also have a map from the steady state ($\d_t =0$) Euler equations with pressure in 2+1 dimensions and the null equations with 
zero presssure, $p=0$, in 3+1 dimensions at $t-z=0$ and $\d_+ v^a=0$. One then finds the Hopfion solution with velocity (\ref{velhopf}) for
$t-z=0$, density $\rho=1$ and pressure
\be
p=p_\infty-\frac{2}{1+x^2+y^2}.
\ee
This is a solution of the incompressible ($\rho=1$ is constant) Euler's equations with pressure in 2+1 dimensions. 

Note however that we have used dimensionless coordinates throughout. 
For a physical solution, as we will need later, we must reintroduce an arbitrary scale $l$ for the solution, such that, for instance,
the 2+1 dimensional fluid solution  is 
\bea
\rho&=&l^{-3}\cr
v_x&=&\frac{2y/l}{1+(x/l)^2+(y/l)^2}\cr
v_y&=&\frac{-2x/l}{1+(x/l)^2+(y/l)^2}\cr
p&=&p_\infty -\frac{2 l^{-3}}{1+(x/l)^2+(y/l)^2}.\label{redfluidhopf}
\eea
Then $1/l$ (or $l_0/l$, where $l_0$ is some fixed scale) acts as a perturbation parameter $\epsilon$: when $\epsilon\rightarrow 0$, we obtain a 
trivial solution.

Similarly, for the null fluid Hopfion solution, we write
\bea
\rho &=& \frac{16 \left((t-z)^2/l^2+(x/l)^2+(y/l)^2+1\right)^2}{\left((t/l)^4-2 (t/l)^2
   \left((x/l)^2+(y/l)^2+(z/l)^2-1\right)+\left((x/l)^2+(y/l)^2+(z/l)^2+1\right)^2\right)^3}\cr
v_x&=&\frac{2(y/l+x(t-z)/l^2)}{1+(x/l)^2+(y/l)^2+(t-z)^2/l^2},\cr 
v_y&=&\frac{-2(x/l-y(t-z)/l^2)}{1+(x/l)^2+(y/l)^2+(t-z)^2/l^2},\cr
v_z^2&=&1-v_x^2-v_y^2.\label{nullfluidhopf}
\eea

\section{Towards a nontrivial helicity in gravity}

We can construct a quantity that has the flavor of the electromagnetic, or the fluid, helicity also in gravity, and argue that it could be made to be conserved
(though we will not find a solution for which it is). 

Consider first a map from electromagnetism in 
a certain static case to gravity. 

In de Donder gauge,
\be
\d^\nu \bar h _{\mu\nu}=0;\;\;\; \bar h_{\mu\nu}\equiv h_{\mu\nu}-\frac{1}{2}\eta_{\mu\nu}h\;,
\ee
the linearized equations of motion for gravity are 
\be
\Box \bar h_{\mu\nu}=-\frac{16\pi G_N}{c^4}T_{\mu\nu}.
\ee
One then considers {\em nonrelativistic} sources such that we have (the theory itself is relativistic!)
\be
\bar h_{00}=\frac{4\Phi}{c^2};\;\;\; \bar h_{0i}=-2\frac{A_i}{c^2};\;\;\; \bar h_{ij}={\cal O}(c^{-4})\;,\label{hbar}
\ee
where we have defined $\Phi$ as the gravitoelectric potential and $A_i$ as the gravitomagnetic potential,
which means that we neglect tensor potentials (giving gravitational waves, for instance). The sources then are the 
"charge density"
\be
\rho=\frac{T^{00}}{c^2}
\ee
and the "current"
\be
j^i=\frac{T^{0i}}{c}.
\ee

Far from these sources, we have 
\be
\Phi\sim \frac{G_N M}{r};\;\;\;\vec{A}\sim \frac{G_N}{c}\frac{\vec{J}\times \vec{r}}{r^3}\;,
\ee
where $M=\int \rho$ and $\vec{J}=\int \vec{j}$ are the total mass and current  (angular momentum) of the source. 
Note that this implies that the gravitomagnetic field of a rotating body is 
\be
\vec{B}_g=\frac{G_N}{2c}\frac{\vec{L}-3\left(\vec{L}\cdot\frac{\vec{r}}{r}\right)\frac{\vec{r}}{r}}{r^3}\sim \frac{1}{r^3}.
\ee

The metric perturbed by the linearized "gravitoelectromagnetic field" is then
\be
ds^2=-c^2\left(1-\frac{2\Phi}{c^2}\right)dt^2-\frac{4}{c}\vec{A}\cdot d\vec{x} dt+\left(1+\frac{2\Phi}{c^2}\right)d\vec{x}^2\/,\label{emgrav}
\ee
meaning that (note that $x^0=ct$) indeed we have (\ref{hbar}).

With the above definitions, we have the "electric" and "magnetic" fields
\bea
\vec{E}&=& -\vec{\nabla}\Phi-\frac{1}{c}\frac{\d }{\d t}\left(\frac{1}{2}\vec{A}\right)\cr
\vec{B}&=& \vec{\nabla}\times \vec{A}\;,
\eea
and the Maxwell equations become
\bea
\vec{\nabla}\cdot \vec{E}=4\pi G_N \rho; && \vec{\nabla}\times \vec{E}=-\frac{1}{c}\frac{\d}{\d t}\frac{\vec{B}}{2}\cr
\vec{\nabla}\cdot \vec{B}=0 && \vec{\nabla}\times \frac{1}{2}\vec{B}=\frac{1}{c}\frac{\d }{\d t}\vec{E}+\frac{4\pi G_N}{c}\vec{j}\;,
\eea
so that $Q_E=M$, but $Q_B=2M$. The de Donder gauge condition gives, for the $\nu =0 $ component, the Lorentz gauge 
condition of electromagnetism,
\be
\frac{1}{c^2}\left[
\frac{1}{c}\d_t \phi+\vec{\nabla}\cdot \left(\frac{1}{2}\vec{A}\right)\right]=0\;,
\ee
in which all terms are of order $1/c^3$ in the nonrelativistic expansion in $1/c$. The $\nu=i$ component on the other 
hand gives at this order
\be
\d_j \bar h_{ji}+\eta^{00}\frac{1}{c}\d_t \bar h_{0i}\simeq +\frac{2}{c^3}\d_t A_i=0\;,
\ee
so $A_i$ is time independent. Note that this is not what is claimed in \cite{Mashhoon:1999nr}, but see the more detailed criticism of that paper's statement
in \cite{FilipeCosta:2006fz}. Thus the map to electromagnetism, despite obtaining the Maxwell equations, is only valid for time-independent $A_i$. 

However, using
the map in (\ref{emgrav}), we see that for instance the magnetic helicity becomes in gravity, at the linearized level defined previously,
\be
H_{mm}=\frac{c^4}{4}\int d^3x \epsilon_{ijk}g_{0i}\d_j g_{0k}.\label{trial}
\ee
But this form is not very satisfactory, as it is not invariant, and moreover $g_{0i}$ is not well defined in general. 

In electromagnetism, in the gauge $A_0=-\phi=0$, possible since we are in vacuum, the helicities are conserved 
on null solutions, $\vec{E}\cdot\vec{B}=\vec{E}^2-\vec{B}^2=0$, like the Hopfion. On the other hand, 
$H_{mm}$ defined above
is conserved in time at the linearized level simply due to the map between gravity and electromagnetism for the 
case $\d_t A_i=0$, though since the Hopfion is time {\em dependent}, we cannot embed it in gravity in this way. 

But we can at least extend the helicity $H_{mm}$ above, to the nonlinear level. At the nonlinear level, consider 
the ADM parametrization of the metric, 
\be
ds^2=-N^2 dt^2+\gamma_{ij}(dx^i+N^i dt)(dx^j+N^j dt)\;,
\ee
which means that
\be
g_{0i}=\gamma_{ij}N^j\equiv N_i;\;\;\; g_{ij}=\gamma_{ij};\;\;\;
g_{00}=-N^2+N^iN_i.
\ee
In summary, the metric and its inverse are
\be
g_{\mu\nu}=\begin{pmatrix}-N^2+N_i N^i & N_i\\ N_i & \gamma_{ij}\end{pmatrix}\;,\;\;\;
g^{\mu\nu}=\begin{pmatrix} -\frac{1}{N^2} & \frac{N^i}{N^2}\\ \frac{N^i}{N^2} & \gamma^{ij}-\frac{N^iN^j}{N^2}\end{pmatrix}.
\ee
Then $N_i\equiv g_{0i}$, $N\equiv \sqrt{-g^{00}}$, $\gamma_{ij}\equiv g_{ij}$. 

Comparing this with the "gravitoelectromagnetic field" form (\ref{emgrav}), 
and keeping only terms linear in the perturbations $\phi$ and $\vec{A}$ and putting $c=1$, we obtain
\bea
\gamma_{ij}&=& (1+2\phi)\delta_{ij}\cr
N_i&=& -2 A_i\Rightarrow N^i\simeq -2 A_i\cr
N_iN^i-N^2&=&-(1-2\phi)\Rightarrow N^2\simeq 1-2\phi.
\eea

In the gauge $\phi=0$, we have $g_{0i}=N_i=N^i\simeq -2 A_i$, so the magnetic helicity can also be written at the 
linearized level, and for $\d_t A_i=0$, as (up to a 
possible multiplicative constant that we ignore)
\be
H_{mm}=\int d^3x \epsilon^{ijk} N_i \d_j N_k.
\ee
But since $N_i$ is well defined at the nonlinear level in the ADM parametrization,
this is the correct nonlinear generalization of our first try (\ref{trial}). 

The condition for it to be conserved (time independent) becomes
\be
\d_t H _{mm}=2\int d^3x \epsilon^{ijk}\dot N_i \d_j N_k=0.
\ee
We can indeed {\em a priori} satisfy this at the nonlinear level, though we will not try to find a solution for which it is. 

In the ADM Hamiltonian formalism, the variables are $\gamma_{ij}$ and their conjugate momenta, whereas 
$N$ and $N_i$ can be arbitrarily given.
Their values can be fixed by gauge invariance, i.e., general coordinate transformations, and by fixing them, 
we obtain the Hamiltonian ($H=0$) and momentum 
($H_i=0$) constraints for their conjugate momenta. In turn, this means that we can choose to have 
a conserved magnetic helicity, $\d_t H_{mm}=0$, at any time (so that $H_{mm}$ is conserved)
for any solution. 

The condition for a nonlinear solution with conserved helicity is thus 
\be
\epsilon^{ijk} \dot N_i \d_j N_k=0\;,\label{null1}
\ee
which is more stringent than the condition for conservation of $H_{mm}$, which would be just its integral to vanish. 
But that is as it should be, since at the linearized level this becomes
\be
\epsilon^{ijk}\dot A_i \d_j A_k=0\Rightarrow \vec{E}\cdot \vec{B}=0\;,
\ee
which is part of the null condition of electromagnetism. Note also that the condition, and the form of $H_{mm}$, are invariant under 
spatial diffeomorphisms, as it should be the case.

There is one more helicity that can be written at the nonlinear level, namely $H_{me}$, given by 
\be
H_{me}=\int d^3x \vec{A}\cdot \vec{E}=-\int d^3x \vec{A}\cdot \dot{\vec{A}}=-\frac{1}{2}\frac{\d}{\d t}\int d^3x \vec{A}^2.
\ee
As we can see, it is written entirely in terms of $\vec{A}$, without making use of $\vec{C}$, which is only implicitly defined 
in the gravity case.

At the linearized level, $A_i$ is replaced by $N_i$, but at the nonlinear level, for invariance under 
spatial diffeomorphism, we need to raise and lower spatial indices by $\gamma_{ij}$.
At the nonlinear level then, the mixed helicity $H_{me}$ becomes either
\be
H_{me}=-\int d^3x  N^i \dot N_i=-\int d^3 x \gamma^{ij} N_i \dot N_j
\ee
or 
\be
\tilde H_{me}=-\frac{1}{2}\frac{\d}{\d t } \int d^3x \gamma^{ij} N_i N_j.
\ee
Moreover, at the nonlinear level we cannot use the Maxwell equations for the condition of conservation of $H_{me}$, 
so we have to impose a condition given by the nonlinear generalizations above, either
\be
\d_t(\gamma^{ij}N_i \dot N_j)=0\;,
\ee
for $H_{me}$, or
\be
\d_t^2(\gamma^{ij} N_i N_j)=0
\ee
for $\tilde H_{me}$.
However, this last possibility is not very plausible, since we would need that nevertheless $\d_t (\gamma^{ij}N_iN_j)\neq 0$, 
which would mean that $N^iN_i=\gamma^{ij}N_i N_j=C t$, where $C$ is a constant.

The first possibility gives instead that 
\be
\dot \gamma^{ij} N_i \dot N_j=-\gamma^{ij}\d_t(N_i \dot N_j)\;,\label{null2}
\ee
which is a combination of an equation of motion, here the time evolution for the variable $\gamma_{ij}$
in ADM parametrization, together with a "null condition,"  which is a condition on $N_i$ in the current case.

\section{A null fluid/gravity correspondence and using it to embed the Hopfion at a black hole horizon}

In this section we aim to embed both the nonrelativistic fluid solution in 2+1 dimensions and the null relativistic fluid solution in 3+1 
dimensions into gravity, via the fluid/gravity correspondence. Since the correspondence was not defined in the null case, we will propose 
a conjecture for it.

\subsection{Review of the usual (nonrelativistic) fluid/gravity correspondence and 2+1 dimensional Hopfion embedding}

The fluid/gravity correspondence (see \cite{Hubeny:2011hd,Rangamani:2009xk,Eling:2010vr} for a review) 
starts with a boosted $AdS_{d+1}$ black hole, for a time-like 4-velocity $u^\mu$, $u^\mu u_\mu=-1$, 
\be
ds^2=-2u_\mu dx^\mu dr -r^2 f(r) u_\mu u_\nu dx^\mu dx^\nu +r^2(\eta_{\mu\nu}+u_\mu u_\nu)dx^\mu dx^\nu\;,\label{boosted}
\ee
where
\be
f(r)=1-\frac{r_H^d}{r^d},\;\;\; r_H=\frac{4\pi }{d}T.
\ee
For $d=4$, we have $r_H=\pi T$ and 
\be
ds^2=-2u_\mu dx^\mu dr+\left(\frac{\pi^4 T^4}{r^2}u_\mu u_\nu+r^2\eta_{\mu\nu}\right)dx^\mu dx^\nu.
\ee
We then promote $u^\mu$ and $T$ to $d$-dimensional fields $u(x^\mu), T(x^\mu)$. This is then not a solution of Einstein's equations anymore, 
so we need to solve Einstein's equations perturbatively, in the gradients of the fields $u(x^\mu)$ and $T(x^\mu)$. 

For a conformal fluid, with fluid energy-momentum tensor
\be
T^{\mu\nu}=\rho u^\mu u^\nu+p(\eta^{\mu\nu}+u^\mu u^\nu)\;,\label{fluid}
\ee
the tracelessness condition gives in the nonrelativistic case with $u^\mu u_\mu=-1$
\be
\eta_{\mu\nu}T^{\mu\nu}=0\Rightarrow \rho=p(d-1).
\ee

In that case, adding the thermodynamic conditions $\rho+p=Ts$ (Gibbs-Duhem) and $dp=sdT$ is enough to 
fix everything as a function of $T(x)$, up to a multiplicative constant, since 
\be
\rho =p(d-1)\Rightarrow p=\frac{sT}{d}\;,
\ee
which together with $dp=sdT$ gives
\be
s=a T^{d-1}\Rightarrow p=\frac{aT^d}{d},\;\; \rho=\frac{(d-1)}{d}aT^d.
\ee
So the conformal fluid is (almost, up the constant $a$) completely determined by $u^\mu(x^\rho)$ and $T(x^\rho)$, including the pressure and energy density. 

But one can also embed the usual Euler (and generalized to Navier-Stokes, and higher corrections in derivatives) 
fluid (see for instance \cite{Eling:2010vr} for a review). By a scaling limit of the conformal case, one obtains, for $u^\mu\simeq 
(1,v^i)$, and keeping only terms of order one and order $v$, 
\be
ds^2=-r^2\, f\, dt^2+ 2dt dr +r^2dx^idx^i-\frac{(4\pi/d T)^d}{r^d} 2v^i dx^idt -2v^i dx^i dt\;,\label{nrfluidd}
\ee
and then turning $v^i$ and $T$ into fields $v^i(t,x^j)$ and $T(t,x^j)$. Moreover, the pressure appears in the 
expansion 
\be
T=T_0(1+\epsilon^2 P(t,x^i))\;,
\ee
and expanding in $\d_i \sim v^i\sim \epsilon$ and $\d_t\sim \epsilon^2$, we obtain the Navier-Stokes equation for 
an {\em incompressible} fluid (meaning that $\rho=$constant, and can be set to 1) $\d_i v^i=0$  with viscosity 
\be
\frac{\eta}{\rho}=\frac{1}{2\pi T_0}. 
\ee
So now the parameter $T_0$ defines the viscosity of the fluid. For $T_0\rightarrow\infty$, we get $\eta/\rho\rightarrow 0$, 
which is the Euler equation (instead of the Navier-Stokes equation). 

Note that the way this happens is as follows: One obtains first the conformal hydrodynamics, with variables $u^\mu(\vec{x},t)$ and $\rho(\vec{x},t)$, 
or rather $T(\vec{x},t)$. After the above scaling limit, we obtain a nonrelativistic theory, with $u^\mu=(1,v^i)$, $v^i\ll 1$, and one has {\em redefined }
variables: $\rho=3(\pi T)^4$ is now considered (approximately) constant, so we have an incompressible fluid, but the {\em small} variation in 
$T$ defines a {\em new variable} $P$, not the pressure $p=\rho/3$ of the conformal fluid, but rather a new pressue, of the nonrelativistic fluid. 

The proof is in the fact that, with this scaling, the conformal fluid equations of motion imply the Navier-Stokes equations of motion, as reviewed in
\cite{Eling:2010vr}. The proof of the fluid gravity correspondence itself, found in  \cite{Bhattacharyya:2008jc}, is that the conditions for the Einstein
equations to be solvable on the ansatz are exactly the conservation of the fluid energy-momentum tensor, i.e., the fluid equations (Navier-Stokes 
in the nonrelativistic case). 

Finally then, we can embed the nonrelativistic fluid Hopfion solution (\ref{redfluidhopf}) into gravity, to leading nontrivial order in $\epsilon=1/l$, 
by using (\ref{nrfluidd}) for $d=3$. Note that the metric (\ref{nrfluidd}) has a proposed gravitational helicity that, at least
in the linearized case, similar to the fluid helicity, 
\be
H_{mm}\simeq \int d^3x \epsilon^{ijk} g_{0i}\d_j g_{0k}\sim \int d^3x\epsilon^{ijk} v_i \d_j v_k =H_v.
\ee
Of course, for the 2+1 dimensional Hopfion this vanishes trivially (since there is no $\epsilon^{ijk}$ now), 
but it makes it likely to find a corresponding 3+1 
dimensional solution that has such $H_{mm}\neq 0$. 

\subsection{Null fluids and gravitational shockwaves}

Other than the standard nonrelativistic limit from the previous subsection, 
there is another way to obtain the Euler equation from the conformal relativistic fluid tensor (\ref{fluid}), namely by considering a null fluid $u^\mu u_\mu=0$ 
with zero pressure, like we have already observed in section 2. Indeed, then the tracelessness of the energy-momentum tensor for $u^\mu u_\mu=0$
implies simply $p=0$. Thus the null fluid solution from section 2 could in principle be embedded in a black hole horizon by the same procedure as 
the one in the previous subsection, except the procedure was defined only for $u^\mu u_\mu=-1$, not for the null case $u^\mu u_\mu=0$. 

That means that we need to generalize to the case of a black hole in AdS space, moving at the speed of light (since the normal procedure started with a 
boosted AdS black hole for $v<c$), with $u^\mu u_\mu=0$. 

But in flat space, boosting a Schwarzschild
black hole to the speed of light results in the Aichelburg-Sexl pp gravitational shockwave, as shown in their original paper 
\cite{Aichelburg:1970dh}. Indeed, they start by considering 
the boosted 4 dimensional Schwarzschild solution in isotropic coordinates, rewritten as 
\be
ds^2=\left(1+\frac{m}{2r}\right)^2 dx^\mu dx^\nu \eta_{\mu\nu}-\left[\left(1+\frac{m}{2r}\right)^4-\left(\frac{1-m/2r}{1+m/2r}\right)^2
\right] dx^\mu u_\mu dx^\nu u_\nu\;,
\ee
where $u_\mu dx^\mu=\gamma(dt'-vdx')$, 
\be
r^2=\gamma(x'-vt')^2+y^2+z^2\;,
\ee
and the limit is taken writing $m=p\sqrt{1-v^2}$ and keeping $p$ fixed as $v\rightarrow 1$. Then, through a somewhat complicated
limiting procedure and change of coordinates, one obtains the A-S shockwave metric
\bea
ds^2&=&dx^\mu dx^\nu \eta_{\mu\nu}-4pG_N\left[\frac{1}{x^+}-\frac{2}{x^-}\delta(x^+)\right]\ln (r^2)(dx^+)^2.\cr
&=&dx^\mu dx^\nu \eta_{\mu\nu}+8pG_N\ln (r^2)\delta(x^+)(dx^+)^2.\label{aichels}
\eea
On the second line, we have performed a coordinate transformation.

The resulting gravitational shockwave is of the general pp wave type. 

A general pp wave in flat spacetime background, in Brinkmann coordinates, is 
\be
ds^2=2dx^+dx^- + H(x^+,x^i)(dx^+)^2+dx_idx_i.
\ee

For pp waves in flat spacetime, the Einstein equation reduces to the single {\em linear} equation
\be
R_{++}=-\frac{1}{2}\d_i^2 H(x^+,x^i)=8\pi G_N T_{++}.
\ee
For the A-S shockwave in flat spacetime, the Einstein equation above
is solved for a particle of momentum $p$ moving at the speed of light, which is the endpoint of the limiting procedure for the black hole 
of mass $m=p\sqrt{1-v^2}$, with energy-momentum tensor
\be
T_{++}=p\delta^{d-2}(x^i)\delta(x^+).
\ee
and gives 
\be
H(x^+,x^i)=\delta(x^+)\Phi(x^i)\;,
\ee
where $\Phi$ satisfies the Poisson equation in $d-2$ dimensions,
\be
\d_i^2\Phi(x^i)=-16\pi G_N p\delta^{d-2}(x^i).
\ee
Note that in the general pp wave case, the Einstein equation is linearized.

The formula on the first line in (\ref{aichels}) is not quite the AS shockwave above, but the extra piece satisfies the vacuum equation, and as we said, the 
pp wave ansatz linearizes the Einstein equations. 

One can put A-S gravitational shockwaves in more general spacetimes, as analyzed in \cite{Nastase:2004pc}.
In particular, the solution inside AdS background was obtained, with ansatz
\be
ds^2=\frac{dx^\mu dx^\nu \eta_{\mu\nu}+dz^2+h(x^+,x^i)(dx^+)^2}{z^2}\;,\label{adsans}
\ee
where
\be
h(x^+,x^i)=\Phi(x^i,z)\delta(x^+).
\ee

The function $\Phi$ is complicated; in $d=4$, it looks somewhat simpler:
\bea
\Phi(x^i,z)&=&p\; 8G_5 R z^4\int_0^\infty J_0(qr) K_2(zRq)I_2(z_0 Rq),\;\;\; {\rm for}\;\;\; z>z_0\cr
&=& 8 G_5 R \int_0^\infty J_0(qr) I_2(zRq) K_2(z_0 Rq),\;\;\; {\rm for}\;\;\; z<z_0.\label{Phi}
\eea
Here $R$ is the scale of $AdS$, $r^2=x^i x^i$,
$z_0$ is the position of the shockwave in the radial direction, and $I_n,J_n,K_n$ are Bessel functions. 
Gubser et al. \cite{Gubser:2008pc} found a similar 
solution, writing an ansatz $\Phi=\tilde \Phi z$, but with a $\tilde \Phi$ written in another way.

\subsection{Conjecture for a null fluid/gravity correspondence}

Since we have a fluid/gravity correspondence for a finite boost by $u^\mu$, we could in principle 
consider the A-S limiting procedure for the 
boosted black hole in AdS space in order to defined the null fluid/gravity correspondence. However,
 that seems technically difficult (already in flat background the procedure of Aichelburg and Sexl is nontrivial), so we can consider
a shortcut instead. 

Just like in flat background we can either consider the Aichelburg-Sexl procedure and do a scaling limit on the boosted black hole to get the shockwave
solution, or directly solve the Einstein equations for the pp wave background with the boosted energy-momentum tensor $T_{++}$, we can 
consider the same second possibility in AdS background. 

Without performing the Aichelburg-Sexl procedure, we can already find that the shockwave that would be obtained must be (equivalent by a 
coordinate transfomation to) the shockwave in AdS background found by simply solving the Einstein equation with the limiting $T_{++}$, i.e., 
for a velocity in the $z$ direction, $v_i=\delta_{iz}$, 
\be
ds^2=\frac{dx^\mu dx^\nu \eta_{\mu\nu}+dz^2}{z^2}+\frac{\Phi(x^i,z)}{z^2}\delta(x^+)(dx^+)^2.
\ee
But note that we can write $x^+=u_\mu x^\mu$, $dx^+=u_\mu dx^\mu$, where $u^\mu=(1,0,0,1)$, so that by rotating to a general frame, 
the solution is
\be
ds^2=\frac{dx^\mu dx^\nu \eta_{\mu\nu}+dz^2}{z^2}+\frac{\Phi(x^i,z)}{z^2}\delta(u_\mu x^\mu)u_\mu u_\nu dx^\mu dx^\nu.
\ee
Here $\Phi(x^i,z)$ has the functional form in (\ref{Phi}), which has $\Phi\propto p$. 

Like in the case of the non-null fluid/gravity correspondence, the next step is to allow for an abitrary, but small, spacetime variations of the 
parameters 4-velocity $u^\mu$ and momentum of the shockwave $p$ (since the 4-momentum is $p^\mu =p u^\mu$), meaning that we consider
now a velocity field $u^\mu=(1,v^i(x^j,t))$ with $v^iv^i=1$, and a momentum field $p=p(x^j,t)$. 
Here $p$ is momentum density, which can be equated with $\rho(x^i,t)$, the energy density of the conformal fluid,
\be
p(\vec{x},t)=\rho(\vec{x},t).
\ee

However, note that unlike the usual, nonrelativistic, fluid-gravity correspondence, we don't need to take a (non-relativistic) scaling to obtain 
the Euler equation. Indeed, as we saw, for the null pressureless conformal case, $p=0$, $u^\mu u_\mu=0$, the conformal fluid equations 
of motion become just Euler. In particular, that means that we don't need a temperature $T$, and to expand it like before, $T=T_0(1+\epsilon^2P)$.
We already have the necessary variables, $\rho(\vec{x},t)$ and $u^\mu(\vec{x},t)$, with $u^\mu u_\mu=0$. 
All we need is that they satisfy the equation of motion 
of a conformal energy-momentum tensor, $\d^\mu T_{\mu\nu}=0$, $T_{\mu\nu}=\rho u_\mu u_\nu$.

Then the natural  zeroth 
order solution for the fluid-gravity correspondence would be
\bea
ds^2&=&\frac{dx^\mu dx^\nu \eta_{\mu\nu}+dz^2}{z^2}+\frac{\phi(r,z)}{z^2} p(\vec{x},t)\delta(u_\mu(\vec{x},t) x^\mu)u_\mu(\vec{x},t) u_\nu(\vec{x},t) 
dx^\mu dx^\nu\cr
&=&\frac{dx^\mu dx^\nu \eta_{\mu\nu}+dz^2}{z^2}+\frac{\phi(r,z)}{z^2} \rho(\vec{x},t)\delta(t-v^i(\vec{x},t)x^i)(dt-v^i(\vec{x},t) dx^i)(dt-v^j(\vec{x},t) dx^j).\cr
&&\label{try1}
\eea
Note that, in effect, we have defined a coordinate $x^+$ by the condition
\be
dx^+\equiv u_\mu (\vec{x},t)dx^\mu\;,
\ee
and there are {\em a priori} two possible choices, differing slightly, for the delta function. One is for its argument to be the one above, and 
the other is for it to be $x^+$, obtained by integrating the above relation (for constant $u^\mu$, the two are identical), i.e.,
\be
ds^2=\frac{dx^\mu dx^\nu \eta_{\mu\nu}+dz^2}{z^2}+\frac{\phi(r,z)}{z^2} \rho(\vec{x},t)\delta(x^+)(dx^+)^2.\label{try2}
\ee

Here $\phi=\Phi/p$ and now $r$ is the radius {\em locally transverse to the shock surface}. 

Like in the non-null case, this is now not a solution of the Einstein equations anymore (in particular, because it describes an arbitrary shock surface 
$v^i(x,t)x^i=t$, moving locally at the speed of light), but solves it only to leading order, and we need to add corrections order by order in derivatives
of the fields. 

But the hope is that, like in the case of the usual fluid-gravity correspondence, there exists  still a solution defined {\em order by order in derivatives}
from a deformation of the above, 
provided the energy-momentum tensor of the dual fluid is conserved {\em order by order in derivatives (in the fluid expansion)}, i.e., provided 
the fluid satisfies the generalized Navier-Stokes (and higher order) equations.

If we would have applied directly the Aichelburg-Sexl limiting procedure on the usual fluid/gravity correspondence, we could have guaranteed that 
the end result was also correct, though even in that case there would have been important questions about the order of limits. As it is, since 
we used our shortcut, the proposal above for a null fluid/gravity correspondence is a conjecture only. 

To test the conjecture, we could in principle do the same procedure  \cite{Bhattacharyya:2008jc} did, and find the equations of motion 
that $\rho(\vec{x},t)$ and $u^\mu(\vec{x},t)$ need to 
satisfy in order for the metric ansatz above to be an approximate solution of the Einstein equations at
leading (first) order in derivatives. In the usual fluid/gravity correspondence case, satisfying the equation of motion at first order in 
derivatives means adding terms to the solution which are linear in derivatives, 
and guaranteeing the conservation of the zeroth order  conformal fluid energy-momentum tensor. 
At this leading order, there will be no difference between the the two proposals, (\ref{try1}) and (\ref{try2}), so we can use either one, as 
one wants. 

However, because now we have a delta function in the ansatz, depending on the fields, the process is more involved, and we have been unable to 
do it so far. The sketch of the procedure is as follows.

The function $\Phi$ was chosen to satisfy 
\be
\frac{1}{z^2}\Delta_{(\vec{x},z)}\Phi(\vec{x},z)=16\pi G_N p\delta^n(x) \delta(z)\Rightarrow \frac{1}{z^2}\Delta_{(\vec{x},z)}\phi(\vec{x},z)=16\pi G_N 
\delta^n(x) \delta(z).
\ee
But now the energy-momentum tensor of the wave is, considering the set-up (\ref{try2}),
\be
T_{++}=\rho(\vec{x},t)\delta(x^+)\delta^n(x)\delta(z)\;,
\ee
and then the Einstein equations in $(x^+,\vec{x},z)$ coordinates are 
\be
R_{++}=8\pi G_N T_{++}
\ee
and the other components zero, just like in the constant case. But now one would need to add new terms, linear in derivatives, to the 
ansatz, to turn it into a first order solution of the Einstein's equations. The claim is that this should (only?) be possible if 
\be
0=\d_\mu T^{\mu\nu}=(\d_\mu \rho) u^\mu u^\nu+\rho u^\nu(\d_\mu u^\mu)+\rho (u^\mu \d_\mu) u^\nu.
\ee

Note that we obtain 
\be
R_{++}=\frac{1}{2}\Delta_{(\vec{x},z)}(\rho \phi \delta (x^+))+{\rm more}\;,
\ee
but now we can't take out $\rho$ and $\delta(x^+)$ from inside the Laplacean, as they depend on $\vec{x}$.

\subsection{Gravitational Hofpion from null fluid Hopfion}

Having defined the null fluid/gravity correspondence, we can just embed the solution (\ref{nullfluidhopf}) into the 
ansatz (\ref{try1}) or (\ref{try2}), viewed as only valid to leading nontrivial order in $\epsilon=1/l\rightarrow 0$.

As we saw in section 3 and at the end of section 4.1, there is a helicity, that can be expressed in terms of gravity 
variables as well. The same will be true in this case. To define it, we first note that the off-diagonal metric is proportional to the momentum of the 
wave, 
\be
g_{0i}\propto p(\vec{x},t)  v^i(\vec{x},t)\equiv p^i(\vec{x},t).
\ee

The fluid Hopfion is characterized by a nonzero fluid helicity
\be
\int d^3x \epsilon^{ijk} v_i\d_j v_k\;,
\ee
which now becomes a nonzero
\be
H=\int d^3x \epsilon^{ijk}g_{0i} \d_j g_{0k}\;,
\ee
which can then be considered as the gravitational analog of the helicity. Indeed, in the case of the linearized gravity, we have seen that 
this corresponds to $H_{mm}$,
\be
H_{mm}=\int d^3x \epsilon^{ijk} N_i \d_j N_k
\ee
in the ADM parametrization, where we also have $g_{0i}=\gamma_{ij} N^j\equiv N_i$, which means we indeed have $H=H_{mm}$.

So we have a nonzero helicity from embedding the fluid Hopfion via the null fluid/gravity correspondence, 
just that now we have a singular case, since
\be
g_{0i}\propto \delta(u_\mu x^\mu).
\ee

\section{Conclusions}

In this paper we have conjectured a null fluid/gravity correspondence, generalizing the usual 
fluid/gravity correspondence to the null case, and used it to 
embed the null fluid Hopfion solution in gravity. We have also trivially embedded the 2+1 dimensional fluid Hopfion in gravity
via the usual (non-null) fluid/gravity correspondence.
We have defined a nonlinear helicity for gravity, an analog of the helicity of electromagnetism (carried by the electromagnetic Hopfion 
solution) and the fluid helicity (carried by the null fluid Hopfion), and have shown that our gravitational null 
Hopfion has a nonzero such helicity, though a very singular one. 

It would be nice to prove explicitly the validity of our conjecture for the null fluid/gravity correspondence, but that seems technically very challenging, so we have left it 
for further work.

\section*{Acknowledgements}
We would like to thank Carlos Hoyos and Jacob Sonnenschein for collaboration at the early stages of this project, and for many
useful comments on the manuscript.
The work of HN is supported in part by CNPq grant 304006/2016-5 and FAPESP grant 2014/18634-9. HN would also 
like to thank the ICTP-SAIFR for their support through FAPESP grant 2016/01343-7, and the University of Cape Town for hospitality during the final stages of this work. 
D.F.W.A. is supported by CNPq grant 146086/2015-5.

\bibliography{HopfionGravity}
\bibliographystyle{utphys}

\end{document}